\documentclass[12pt]{article}
\usepackage{amsmath}
\usepackage[latin1]{inputenc}
\usepackage{graphicx}
\usepackage{float}
\usepackage{amsfonts}
\usepackage{hyperref}
\usepackage{graphicx}
\usepackage{multimedia}
\usepackage[mathscr]{eucal} 

\begin{document}
\title{Hybrid mesons as systems of confined monopoles}
\author{~L.~E.~Oxman\\ \\
Instituto de F\'{\i}sica, Universidade Federal Fluminense,\\
Campus da Praia Vermelha, Niter\'oi, 24210-340, RJ, Brazil.}
\date{\today}
\maketitle
 
\begin{abstract}
We review some recent ideas regarding classical topological objects in dual superconductor models that could represent different confining states of the gluon field. We also comment about natural components in (magnetic) ensembles that could effectively originate these models at large distances. 
 
\end{abstract}
 
\section{Introduction}

The nonperturbative low energy regime of Quantum Chromodynamics (QCD) has been successfully  tested by   ab-initio calculations with dynamical quarks, that permitted to make contact with hadron experimental masses \cite{durr}. 
Searching for an understanding of the mechanism of confinement, physicists were also led to analyze pure Yang-Mills (YM) theories in the presence of nondynamical (heavy) quark-antiquark chromoelectric sources. In particular, many lattice studies have been oriented to obtain the static interquark potential, provided by the Wilson loop average over the pure YM degrees of freedom. Among the properties observed in the lattice, we have,

\begin{itemize}  

\item Asymptotic linearity \cite{Bali}

\item String-like behavior \cite{LW}

\item $N$-ality at asymptotic distances \cite{dFK}:
string tensions depend on how the center is realized in a given $SU(N)$ quark representation. In particular, an adjoint quark source can be  bound to a gluon and form a colourless state.  
This means that the string between $q $, $\bar{q}$ adjoint sources is broken at large distances, when the string energy attains  twice the mass of a quark-gluon state.  

\end{itemize}

Dual superconductivity \cite{N}-\cite{3} offers an important scenario where a confining stringlike behaviour could be explained. The general idea is that (quantum) chromomagnetic degrees of freedom could capture the path integral measure in pure YM. These degrees could condense, giving rise to a dual superconductor of chromomagnetic charges, that confines chromoelectric charges. \vspace{.2in}

This scenario has been explored by means of:

\begin{itemize}

\item Lattice calculations: The path integral measure $[DA_\mu]$ could be captured by quantum ensembles of magnetic center vortices, magnetic monopoles, and correlated combinations of both. Scenarios only based on Abelian projected monopoles are not good at describing $N$-ality. This is in contrast with the situation when center vortex degrees of freedom are present. See \cite{ref4}-\cite{ref19}, and refs. therein.

\item Effective dual models in a Higgs phase: Here, there are phenomenological dimensionful scales from the beginning. In the dual superconductor, the confining string is a smooth vortex solution to the classical equations of motion.  
This is a magnetic object in the dual theory, that is supposed to effectively describe the electric confining string.
In this context, $N$-ality would be naturally implemented if this vortex were a {\it center} vortex, which should not to be confused with the center vortex quantum magnetic degrees of freedom considered in pure YM simulations. See \cite{Baker1}-\cite{notes}, and refs. therein.

\end{itemize}

Here, we review some recent theoretical work we have done about effective models, their possible confining states, and underlying ensembles.

\section{Confining string as a classical dual center vortex}

Effective models are naturally constructed by implementing the SSB pattern that leads to center vortex classical solutions, thus incorporating $N$-ality. For this objective, a model with a Higgs potential that leads to $SU(N) \to Z(N)$ SSB, containing up to quartic terms, must be considered. For this pattern, at least $N$ adjoint Higgs fields are required \cite{Fidel2}. A point of contact between these models and the underlying theory they are supposed to describe is provided by the comparision of the pure YM lattice potential and the energy of the classical dual center vortex. For $SU(3)$, This was done  using a dual model with three adjoint Higgs fields, and a Higgs potential defined  on an ansatz along off-diagonal directions \cite{Baker1,Baker3}.
In the next section, we call the attention to another important point of contact, namely, the possibility of describing hybrid states where a colour nonsinglet $q\bar{q}'$ and a valence gluon form a colourless hybrid meson. 

Before proceeding, we quote a natural class of Yang-Mills-Higgs (YMH) $(3+1)D$ effective models with $SU(N)\to Z(N)$ SSB \cite{ref28}. They depend on a gauge field $\Lambda_\mu$ and a set of $SU(N)$ adjoint Higgs fields,  $\psi_I \in \mathfrak{su}(N)$,
\[
{\cal L} = \frac{1}{2} \langle D_\mu \psi_I , D^\mu \psi_I\rangle +
\frac{1}{4} \langle F_{\mu \nu}, F^{\mu \nu}\rangle - V_{\rm Higgs}(\psi_I)\;,
\]
\[
D_\mu=\partial_\mu-ig [\Lambda_\mu, ~]\makebox[.5in]{,}
F_{\mu \nu}= \partial_\mu \Lambda_\nu -\partial_\nu \Lambda_\mu -i g
[\Lambda_\mu,  \Lambda_\nu]  \;.
\]
Here, $I$ is a flavour index ($I=1, \dots, d$, $d\geq N$). The Higgs potential   
is constructed with the natural $SU(N)$ invariant terms,
\[
\langle \psi_I,\psi_J \rangle 
\makebox[.3in]{,} \langle\psi_I,\psi_J \wedge \psi_K\rangle
\makebox[.3in]{,}
\langle \psi_I\wedge \psi_J,\psi_K \wedge \psi_L\rangle
\makebox[.3in]{,}
\langle \psi_I,\psi_J \rangle \langle \psi_K,\psi_L \rangle \;,
\]
 \[
\psi_I \wedge \psi_J = -i [\psi_I,\psi_J]\;.
\]

\subsection{ Example: flavour symmetric model}

To motivate the construction, we recall that followed for a model with a single $SU(2)$ adjoint   Higgs field $\psi$ undergoing $SU(2)\to U(1) $ SSB. This pattern is obtained from a Higgs potential whose vacua are points on $S^2$,
\[
\langle \psi_0 , \psi_0 \rangle - v^2 =0 \;,
\]
as a given vacuum is left invariant under a $U(1)$ subgroup. The natural Higgs potential, with up to quartic terms, is obtained by squaring the vacua condition,
\[
V_{\rm Higgs}=  \frac{\lambda}{4} \left( \langle \psi , \psi \rangle- v^2\right)^2 \;.
\] 

Now, to get a flavour symmetric model with $SU(N) \to Z(N)$ SSB, we take $d=N^2-1$, so that the range of the flavour index coincides with that of colour. Replacing
 $I \to A=1,\dots, N^2-1$, we denote the Higgs fields as
$\psi_A$ and initially propose a Higgs potential whose vacua satisfy,
\[
\psi_A^0 \wedge \psi_B^0 -  v_c\, f_{ABC}\, \psi^0_C =0\;,
\]
where $f_{ABC}$ are structure constants of $\mathfrak{su}(N)$. The solutions correspond to a trivial point 
$\psi_A^0 =0$ and a manifold of nontrivial points, where  $\psi_A^0$ form a Lie basis.
Of course, the set of solutions is invariant under the adjoint action of $SU(N)$ gauge transformations 
$\psi_A^0 \to U \psi_A^0 U^{-1}$. In addition, a given nontrivial point $(\psi_1^0, \dots , \psi_{N^2-1}^0)$ is invariant
under this action iff $U\in Z(N)$. Then, a natural potential would be obtained by squaring the condition above,
\[
V_{{\rm Higgs}}  =  \frac{\lambda}{4}\,\langle \psi_{A} \wedge \psi_{B} -f_{ABC}\, v_{c}\,\psi_{C}\rangle^{2} \;.
\]
However, for this potential the trivial and nontrivial vacua are degenerate. This can be avoided by initially  expanding the squares and then introducing general couplings for the quadratic, cubic an quartic terms \cite{ref28},
\begin{eqnarray}  
&& V_{\rm Higgs} = c+ \frac{m^2}{2}\, \langle \psi_A ,\psi_A \rangle   +\frac{\gamma}{3} \,f_{ABC} \langle \psi_A \wedge \psi_B,\psi_C \rangle
+ \frac{\lambda}{4}\, \langle \psi_A \wedge \psi_B,\psi_A\wedge \psi_B \rangle  \;.
\nonumber  
\end{eqnarray}
 At $m^{2}=\frac{2}{9}\frac{\gamma^{2}}{\lambda}$ we reobtain the degenerate case, while for $m^2 < \frac{2}{9}\frac{\gamma^{2}}{\lambda}$ the absolute minima are only given by nontrivial vacua,
\[
\psi_A^0=  v_c\, S T_A S^ {-1}  \makebox[.5in]{,}
v_c =-\frac{\kappa}{2\lambda}\pm \sqrt{\left(\frac{\kappa}{2\lambda}\right)^2-\frac{\mu^2}{\lambda }}\;.
\]
Besides gauge invariance, this potential is flavour symmetric under $Ad(SU(N))$ global transformations, $\psi_A \to R_{AB} \, \psi_B$ .


 \section{Hybrid states }

In addition to normal mesons, lattice calculations predict a rich spectrum of exotic objects.
Some of them correspond to  $qg\bar{q}'$ hybrid mesons 
where a nonsinglet colour pair and a valence gluon form a colourless state.  For a review, see ref. \cite{hybrids}. 
This year, a collaboration based at the Jefferson Lab (GlueX) will start mapping gluonic excitations by searching hybrid  states generated by photoproduction. In a world of heavy quarks, a successful effective model should acomodate these hybrid excitations and, besides the normal $Q\bar{q}$ potential, it should also reproduce the lattice hybrid potentials \cite{JKM}. Now, if normal confining strings are to be seen as center vortices, valence gluons should be monopole-like objects interpolating them (colour adaptors) (see ref. \cite{ref28}).  
Denoting the manifold of absolute minima by $ {\cal M}$, when a gauge model undergoes $SU(N) \to Z(N)$ SSB, the manifold is the coset ${\cal M}= SU(N)/Z(N) = Ad(SU(N))$ (the adjoint representation of $SU(N)$). \vspace{.3cm}

The mathematical consequences are: 
\vspace{.2cm}
\begin{itemize}

\item[1)]  Due to $\Pi_1 (Ad(SU(N))) = Z(N) $, there are smooth center vortices.  

\item[2)]  As a compact group has trivial $\Pi_2$, then   $\Pi_2({\cal M})=\Pi_2 (Ad(SU(N))) = 0$, and there are no isolated monopoles. 

\item[3)] There is an exact homotopy sequence  that leads to junctions formed by different center vortices interpolated by a non Abelian monopole. 

\end{itemize}

\vspace{.1cm}

The corresponding physical consequences are:
\vspace{.2cm}
\begin{itemize}

\item[i)]  {\bf There are strings that confine colourless quark states, forming normal hadrons}.

\item[ii)]
{\bf  (valence) Gluons are confined}.

\item[iii)]
{\bf  There are junctions where strings with different colours are interpolated by a valence gluon. They confine  nonsinglet quark states, forming hybrids}.

\end{itemize}

In order to understand the previous statements, it is useful to introduce a polar decomposition for a tuple of adjoint Higgs fields $(\psi_1, \dots, \psi_d)$, $d\geq N$,   in terms of  ``modulus'' $q_I$ and ``phase'' $S$ variables, $\psi_I = S q_I S^{-1}$. This is similar to the procedure used for Nielsen-Olesen vortices, where the complex field is parameterized by $\phi = h \, e^{i\chi}$. Taking $\chi$ as the polar angle, and $h$ interpolating the true asymptotic vacua and the false 
vacuum at the vortex guiding center, smooth finite energy solutions are obtained.  For three $SU(2)$ adjoint fields $\psi_1$, $\psi_2$, $\psi_3$ the usual polar decomposition of a $3\times 3$ matrix can be used. This is done in terms of a symmetric positive semidefinite matrix times a matrix in $SO(3)$, which coincides with the adjoint representation of $SU(2)$. 
For general $SU(N)$, see ref. \cite{OS}.
The non Abelian phase variables $S$ can also be visualized in terms of a local frame in colour space $n_A = S T_A S^{-1}$ (Expanding $q_I $ in a Lie basis). 

\section{Normal glue}

$Z(N)$ center vortices can be labelled by the weights of the different group representations \cite{Konishi-Spanu},
\[
S= e^{i\varphi\, \vec{\beta} \cdot \vec{T}}   \makebox[.5in]{,} \vec{\beta} = 2N\, \vec{w} \;.
\]
A weight $\vec{w}$ is defined by the eigenvalues of diagonal generators corresponding to one common eigenvector,
\[
[T_q,T_p]=0  \makebox[.5in]{,} T_q\,{\rm  eigenvector} = \vec{w}|_q\, {\rm  eigenvector} \;.
\]
For the fundamental representation we have $N$ weights (fundamental colours),
$\vec{\beta}_1 + \dots +\vec{\beta}_N = \vec{0}$. They are associated with the simplest center vortices
\[
e^{i2\pi \, \vec{\beta}_i \cdot \vec{T}} = e^{i\, 2\pi/N}\, I \;.
\]
For this phase $S$, some frame components contain defects so that the modulus variables $q_I$ must satisfy appropriate boundary conditions to render the energy per unit length finite.
For $SU(3)$ we have three weights $\vec{\beta}_1, \vec{\beta}_2, \vec{\beta}_3$ ($\vec{\beta}_{1} + \vec{\beta}_{2} + \vec{\beta}_{3} = 0$).
They represent three possible colours for the fundamental confining string, 
\vspace{.5cm}

\begin{center}
\includegraphics[scale=.25]{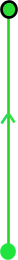} 
\hspace{2cm} 
\includegraphics[scale=.25]{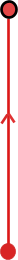} 
\hspace{2cm}
\includegraphics[scale=.25]{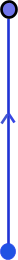} 
\end{center} 
\vspace{.2cm} 
For example, the non Abelian phase that  can be used to confine three quarks to form a baryon in a $Y$-junction configuration
is (see Fig. A1),
\begin{equation}
S= e^{i\chi_1\, \vec{\beta}_1 \cdot \vec{T}} \, e^{i\chi_2\, \vec{\beta}_2 \cdot \vec{T}}  \;,
\label{SY}
\end{equation}
where $\chi_1$ and $\chi_2$ are multivalued when we go around a pair of curves ${\cal C}_1$ and ${\cal C}_2$. These curves coincide on a branch, so that if a center vortex with charge $\vec{\beta}_1$ (resp. $\vec{\beta}_2$) is leaving a pair of  monopoles (representing the green and red quarks, respectively),
then a flux $\vec{\beta}_1 + \vec{\beta}_2$ will enter the third monopole. That is, the third monopole charge is 
$-\vec{\beta}_1 - \vec{\beta}_2 =\vec{\beta}_3$, thus representing the blue quark. 

\vspace{.3cm}

\includegraphics[scale=.32]{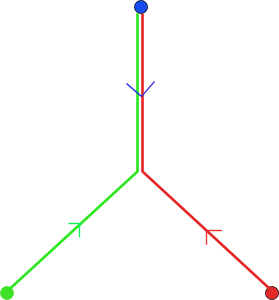} 
~A1
\hspace{1.5cm} 
\includegraphics[scale=.27]{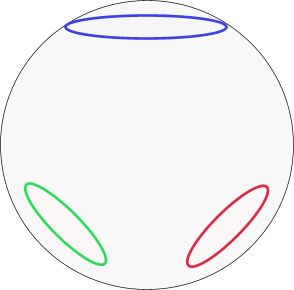}  
~A2
\hspace{1.5cm} 
\includegraphics[scale=.29]{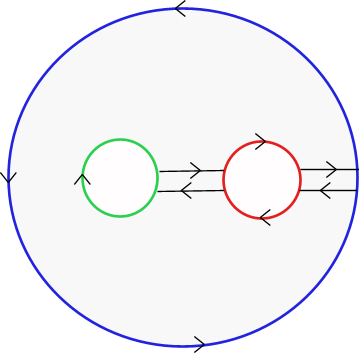}
~A3
\vspace{.5cm}
 
In Fig. A2, we show the weight that dominates the local frame behaviour when we go close and around the three center vortices in a $Y$-junction In Fig. A3, we show this information on a disk whose border is obtained by expanding the small circle around the north pole in Fig. A3. 

\subsection{Hybrid glue}

From the $S$ mapping in eq. (\ref{SY}) it is clear that in every point inside the region R shown in Fig. A3 (region in gray), formed by the disk minus a pair of holes, there are no frame singularities. This occurs in spite of the fact that some frame components are transformed by a Cartan rotation with weights $\vec{\beta}_1$, $\vec{\beta}_2$ and $\vec{\beta}_3$ when the green, red and blue circles are followed.
Another way of looking at this is noting that the map ${\cal C }\to Ad(SU(3)) $, given by $Ad(S)$ defined on the composition of the three coloured circles (oriented as in Fig. A3) and the additional segments in Fig. A3, is a topologically trivial loop in $Ad(SU(3))$. If we start with $S = I$, at the end of the circuit we have,
\begin{equation}
e^{i2\pi \, \vec{\beta}_1 \cdot \vec{T}} e^{i2\pi \, \vec{\beta}_2 \cdot \vec{T}}  e^{i2\pi \, \vec{\beta}_3 \cdot \vec{T}}   = I \;.
\label{m1}
\end{equation}
Then, the corresponding path ${\cal C }\to SU(3) $ is closed and, as $SU(3)$ is simply connected, it can be deformed to a point in R, without further singularities. 

This discussion motivates a different situation involving a pair of frame singularities on the sphere $S^2$ around a point,  which is not equivalent to a normal center vortex. Let us consider a mapping that on a disk minus a hole is defined as shown in Fig. B1.  

\vspace{.5cm}
\includegraphics[scale=.3]{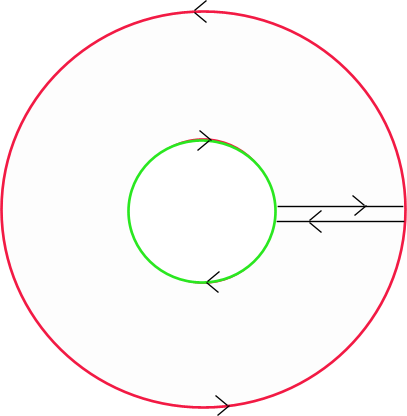}
~B1
\hspace{1.5cm}
\includegraphics[scale=.3]{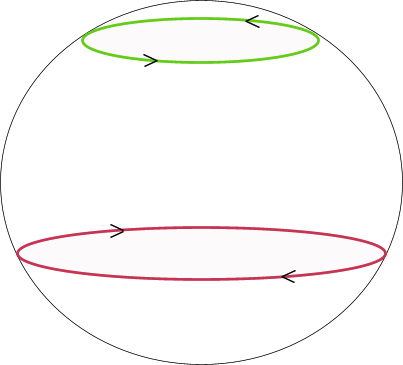}  
~B2
\hspace{2cm}
\includegraphics[scale=.35]{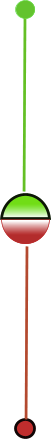} 
~B3
\vspace{.5cm}

When the green and red circles (oriented as in the figure) are followed, the frame is rotated with $\vec{\beta}_1$ and $-\vec{\beta}_2$, respectively. We note that in principle there are no obstructions to extended this behaviour inside the grey region. In effect, the loop ${\cal C }\to Ad(SU(3)) $ defined on the composition of the two coloured circles and the additional segments in Fig. B1 is topologically trivial in $Ad(SU(3))$. Again, it is associated with a closed loop in $SU(3)$, as the final point will be,
\begin{equation}
e^{i2\pi \, \vec{\beta}_1 \cdot \vec{T}} \, e^{-i2\pi \, \vec{\beta}_2 \cdot \vec{T}}  = I \;.
\label{m2}
\end{equation}
Associating the behaviour on the green (red) circle with a defect at the north (south) pole, the fluxes with respect to the $\hat{z}$-axis become $\vec{\beta}_1$ and $\vec{\beta}_2$, respectively. Then, we expect a field configuration where (fundamental, external) red and antigreen monopoles  are bound by center vortices with different weights, interpolated by some pointlike object. Indeed, the mapping required to describe this situation is \cite{ref28}, 
\[
S= e^{i\varphi\, \vec{\beta}_1 \cdot \vec{T}}\, W(x)
\makebox[.5in]{,}
W(x)=e^{i\theta \,\sqrt{N} T_{\alpha} }  \;.
\]
Around the north pole, 
\[
S(x) \sim  e^{i\varphi\, \vec{\beta}_1 \cdot \vec{T}} \;.
\]
Close to the south pole, $W(x)\sim W_{\alpha}=e^{i\pi \,\sqrt{N} T_{\alpha}}$ becomes a Weyl reflection,
so for $\vec{\alpha}=\vec{w}_{1}-\vec{w}_2$, we get the behaviour,

\[
S(x) \sim W_{\alpha}\, e^{i\varphi\, \vec{\beta}_2 \cdot \vec{T}} \;,
\]
and the charge of the interpolating monopole is,
\[
\vec{Q}_m=\frac{2\pi}{g}\, (  \vec{\beta}_{1}   - \vec{\beta}_{2}  ) = \frac{2\pi}{g}\, 2N\, \vec{\alpha} \;.
\]
As the roots are the weights of the adjoint representation, which acts via commutators,
\[
[T_q, E_\alpha]= \vec{\alpha}|_q\, E_\alpha \;,
\] 
this monopole is naturally identified with a confined valence gluon with adjoint colour $\vec{\alpha}$. 

\section{Some topological considerations}

For general $SU(N)$, the (adaptor) monopole part of the configuration is due to pointlike defects in the local diagonal frame components, \[n_q=ST_q S^ {-1} \makebox[.5in]{,} 
q=1, \dots, N-1 \;. \]
These components do not see the vortex like defects in $S$, so they are smooth on a sphere $S^2$ around the center of the monopole.  The set of $n_q$'s can be identified with the quotient space $Ad(G)/Ad(H)$

\[G=SU(N) \makebox[.5in]{,} H=U(1)^{N-1}  \;. \]
The exact sequence,
\[
\Pi_2(Ad(G))\, \overset{f_1}{\rightarrow} \, \Pi_2 \left(\frac{Ad\, (G)}{Ad\, (H)}\right) \,\overset{f_2}{\rightarrow}\, \Pi_1(Ad\, (H)) \,\overset{f_3}{\rightarrow} \, \Pi_1(Ad\, (G)) \;, 
\]
where $Im(f_1)=Ker(f_2)$,  $Im(f_2)=Ker(f_3)$, is behind the topological stability of the monopole junction.
As the second homotopy group of a compact group is trivial, we have,
\[
\Pi_2(Ad(G)) =0 ~ \Rightarrow ~ Im(f_1) = 0 ~ \Rightarrow ~ Ker(f_2) = 0 \;, 
\]
that is, $f_2$ is injective, and there is a one to one mapping between $\Pi_2 \left(\frac{Ad\, (G)}{Ad\, (H)}\right)$ and $Im(f_2)$. On the other hand, $Im(f_2)= Ker(f_3)$ = loops in $\Pi_1(Ad\, (H))$ that are trivial when seen as loops in $\Pi_1(Ad\, (G))=Z(N)$ . In other words,  homotopy classes of $n_q$'s are in one to one correspondence with classes of closed paths in $\Pi_1(Ad\, (H))$ that are also closed in $G=SU(N)$ \cite{ref28}. The latter requirement is given by eqs. (\ref{m1}) and (\ref{m2}), and the classes in $\Pi_1(Ad\, (H))$  are represented by  loops defined on ${\cal C}  \sim S^1$, obtained by the composition of transformations along the coloured circuits. 

For more information about nonabelian vortices, monopoles, as well as the use of exact sequences to analyze complexes formed by them, in supersymmetric theories, see refs. \cite{HD,Dav}, \cite{notes}.

\section{Phenomenological ensembles}

In the introduction, we commented about two different approaches to dual superconductors. One based on lattice calculations, aimed at detecting the ensemble of quantum magnetic degrees of freedom that could capture the path integral measure in pure Yang-Mills theories. The other proposes an effective 
model in a Higgs phase, with phenomenological parameters, relying on general principles and symmetries.
This is analogous to what happens in BCS superconductors. They are understood in terms of microscopic degrees, where the electron-phonon interaction leads to Cooper pairs that condense, as well as via the Guinzburg-Landau model, that captures the main physics of the condensate. From this point of view, 
we may wonder what is the ensemble of magnetic degrees underlying the effective models discussed in previous sections. This type of question has been extensively considered in other contexts, ranging from  compact QED in three and four dimensions \cite{polya} to polymer field theory \cite{ref29}. 

From the beginning, it is easy to advance that adjoint Higgs fields should be obtained as an effective description of pointlike magnetic degrees carrying adjoint charges. The quantitative relationship between both phenomenological models, namely, the effective description and the corresponding ensemble, was initiated  in \cite{OST}. There, we considered an ensemble of looplike monopoles in $4D$. 
The simplest properties to characterize a loop are their length and curvature,  that are physically manifested through a tension $\mu$ and stiffness $1/\kappa$. The coupling of a coloured loop to a non Abelian gauge field was introduced in \cite{ref26},
\[
 u_\mu(s)  \, I^{A}\Lambda_{\mu}^{A}(x(s)) \makebox[.5in]{,} I^{A}=T^{A}_{cd} \,
\bar{z}_{c} z_{d} \makebox[.5in]{,}  u_\mu = \dot{x}_\mu  \;,
\]
where the index $a$ ranges from 1 to $ \mathscr{D}$, the dimension of the group representation.
Now, consider the ensemble,
\[
Z=\int [D\phi]\, e^{-W}\, \sum_n\, Z_n  \;,
\]
where $n$ sums over the number of loops,
\[
Z_n= \int [Dm]_n \, \exp \left[- S^0 +  {\sum_{k=1}^{n}\oint_{L_{k}} ds\;
\left(ig\,  \dot{x}^{(k)}_\mu I^{A}\Lambda_{\mu}^{A}(x^{(k)})  -\phi(x^{(k)})\right) }\right]  \;,
\]
\[
S^0=\sum_{k=1}^{n}{\oint_{L_{k}} ds\, \biggl[ \mu + \frac{1}{2} (\bar{z}_c
\dot{z}_c - \dot{\bar{z}}_c z_c)+ \frac{1}{2\kappa}\, \dot{u}^{(k)}_\mu
\dot{u}^{(k)}_\mu  \biggr]} \;.
\] 
$W$ encodes the loop interaction. In particular, excluded volume effects are implemented with a $\phi^2$-term in $W$.  \vspace{.5cm}

\includegraphics[scale=.3]{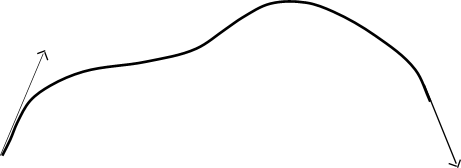} 
\hspace{2cm}
\includegraphics[scale=.3]{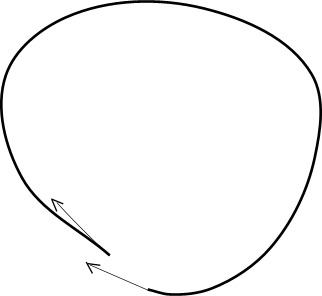}   

\vspace{.3cm}
Fig. C: From an open curve weight to a smooth loop weight. 

\vspace{.3cm}

\noindent
For smooth loops, the partition function can be rewritten as, 
 \[
Z =\int [D\phi]\, e^{-W}\,
e^{\int_{0}^{\infty}\frac{dL}{L}\; \int_{\Re^{4}} d^{4}x \,
\sum_{a} Q^{aa}(x,x,L)}  \;,
\]
\[
Q^ {aa}(x,x,L) =\int d^3u\;   Q^{aa}(x,x,u,u,L) \;. 
\]
$Q^{ba}(x,x_0,u,u_0,L)$ is the
end-to-end probability, for a line of length $L$, to start at $x_0$, with tangent $u_0$ and colour $a$, and end at $x$ with $u$, $b$. 
 The open curve weight has a path-integral representation
that can be obtained as the continuum limit of a polymer growth process \cite{ref24}. This is controled by a Chapman-Kolmogorov recurrence relation for difussion in $x$ and in tangent $u$-space, that leads to the Fokker-Plank equation \cite{OST},
\[
\left[(\partial_L - (\kappa /\pi)\, \hat{L}^{2}_{u}+(\mu +\phi )\,
1 + u\cdot D\right] Q(x,x_0,u, u_{0},L)=0 \;,
\]
\begin{eqnarray}
Q(x,x_0,u,u_0,0)= \delta(x-x_{0})\, \delta(u-u_{0})\, 1\;, \nonumber
\label{iniab1}
\end{eqnarray}
where $Q|^{cd} =Q^{cd}$, $1$ is a $\mathscr{D} \times \mathscr{D}$ identity matrix, and
\[
D_\mu = 1\, \partial_{\mu} - ig\Lambda_{\mu}^{A}T^{A} \;.
\]

\vspace{.5cm}
\includegraphics[scale=.3]{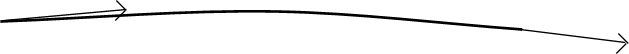} 
\hspace{1.5cm}
\includegraphics[scale=.3]{poly-a.png}
\vspace{.3cm}

Fig. D:  For smaller $1/\kappa$, less $l$-values are needed  to describe the final $u$-distribution. 

\vspace{.5cm}

\noindent
In the semiflexible limit, we can disregard the angular momenta $l\geq 2$  in an expansion of spherical harmonics on $S^3$ (memory loss), obtaining,
\vspace{.2cm}
\[
{\cal Q} (x,x_0,L) \approx \langle x|e^{-L O}  |x_0\rangle \]
\[
O = - \frac{\pi}{12 \kappa}\, D_\mu D_\mu + (\phi +\mu ) \, 1 \;,
\]
and the effective field representation,
\begin{eqnarray}
Z &=& \int [D\phi]\, e^{-W}\, ({\rm Det }\, O)^{-1} \nonumber \\
&=&
\int [D\phi]\, e^{-W} \int [D\zeta][D\bar{\zeta}]\; e^{-\int
d^{4}x\, {\cal L}(\zeta,\Lambda,\phi) } \;.  \nonumber
\end{eqnarray}
When an ensemble of loops with adjoint charges is considered, with a loop interaction obtained by replacing $\phi(x) \to \phi(x)   +I^{A}\Phi^{A}(x)$ and integrating with an appropriate Gaussian weight, ${\cal L}$ can be written in terms of a pair of Hermitian adjoint Higgs fields,
\begin{eqnarray}
\lefteqn{ {\cal L}_{\rm eff}(\psi,\Lambda)= \frac{1}{2} \langle D_\mu \psi_I ,
D^\mu \psi_I\rangle  } \nonumber \\
&& + \frac{m^2}{2} \langle \psi_I,\psi_I \rangle + \frac{\lambda}{4}\, \langle
\psi_I\wedge \psi_J,\psi_I \wedge \psi_J\rangle  + \frac{\eta}{4}\, \langle
\psi_I,\psi_I \rangle \langle \psi_J,\psi_J \rangle \;, \nonumber
\end{eqnarray}
where  $m^2 \propto \mu \, \kappa$. These are those terms involving a pair of flavours in our 
previously discussed effective model.

\section{Discussion}

The detailed knowledge about hadron states and interquark potentials observed in the lattice, for different groups and representations, should guide the search for the natural dual superconductor model. How to accomodate Casimir scaling at intermediate distances, asymptotic $N$-ality, and the interquark potential in normal and hybrid states? 

We believe that hybrid  states, with their intrinsic non Abelian features, are ideal to guide this quest. Confined valence gluons can be identified with confined non Abelian monopoles, that can interpolate a pair of strings with different colours. Note that the experimental search for hybrid states will start 
running this year at the GlueX facility. 

Another important point is that in order to describe  $N$-ality, the dual superconductor must be based on a set of adjoint Higgs fields. As explained in \S 6, this type of model could be an effective description of  {\it quantum} ensembles containing monopole degrees of freedom that carry adjoint $SU(N)$ charges. Thus, while Abelian projected monopoles are not good at desribing $N$-ality, the non Abelian degrees could be an alternative/complementary source to ensembles of {\it quantum} center vortices  and their ensuing properties.

\section*{Acknowledgements}

I would like to thank the organizers of the Fourth Workshop on Nonperturbative Quantum Field Theory, Sophia-Antipolis, France,  for the nice atmosphere and warm hospitality.  The Conselho Nacional de Desenvolvimento Cient\'{\i}fico e Tecnol\'{o}gico (CNPq), the Coordena\c c\~ao de Aperfei\c coamento de Pessoal de N\'{\i}vel Superior (CAPES), and FAPERJ are acknowledged for financial support.

\end{document}